\documentclass[11pt,aps,prd,preprint,nofootinbib]{revtex4}
\usepackage{graphicx,amsmath,amssymb}
\usepackage[usenames,dvipsnames]{pstricks}
\usepackage{epsfig}

\begin{document}

\title{The Theory of (Exclusively) Local Beables}
\author{Travis Norsen}
\affiliation{Marlboro College \\ Marlboro, VT  05344
\\
\vspace{.5 cm}
}

\date{June 17, 2010
\\
\vspace{1 cm}}

\begin{abstract}
It is shown how, starting with the de Broglie - Bohm pilot-wave
theory, one can construct a
new theory of the sort envisioned by several of QM's founders:  a
Theory of Exclusively Local Beables (TELB).  In particular, the usual 
quantum mechanical wave function (a function on a high-dimensional
configuration space) is not among the beables posited by the new
theory.  Instead, each particle has an associated ``pilot-wave'' field
(living in physical space).  A number of additional 
fields (also fields on physical space) maintain
what is described, in ordinary quantum theory, as ``entanglement.''
The theory allows some interesting new perspective on the kind of
causation involved in pilot-wave theories in general.  And it provides
also a concrete example of an empirically viable quantum theory
in whose formulation the wave function (on configuration space) does
not appear  -- i.e., it is a theory according to which nothing 
corresponding to the configuration space wave function need actually 
exist.  That is the theory's \emph{raison d'etre} and perhaps its
only virtue.  Its vices include the fact that it only reproduces the 
empirical predictions of the ordinary pilot-wave theory (equivalent,
of course, to the predictions of ordinary quantum theory) for spinless
non-relativistic particles, and only then for wave functions that are
everywhere analytic.  The goal is thus not to recommend the TELB
proposed here as a replacement for ordinary pilot-wave theory (or
ordinary quantum theory), but is rather to illustrate (with a crude
first stab) that it might be possible to construct a plausible,
empirically viable TELB, and to recommend this as an interesting and
perhaps-fruitful program for future research.  
\end{abstract}

\maketitle

\section{Introduction}

From the very beginning, the quantum revolution centered
around the idea of ``wave-particle duality''.  Einstein's
revolution-triggering 1905 paper (that is, the one titled
``Concerning an heuristic point 
of view toward the emission and transformation of light'') begins 
with a discussion of the ``profound formal distinction'' between the
continuous fields (described by Maxwell's theory of electromagnetic 
processes) and discontinuous particles (exemplified by atoms and
electrons) of classical physics. \cite{einstein}
The need for a novel theory of course arose from the appearance, in
certain key experiments, of discontinuous (particle-like) 
properties in light -- and (later) continuous (field-like) properties in
electrons and other material particles.  

The obvious and natural way of accounting for such ``dual''
appearances is
simply to take the duality literally -- that is, to say that 
what we call a ``photon'' or ``electron''
actually comprises two distinct (though inseparable and interacting)
entities:  a point-like particle, and an associated wave which somehow
guides or choreographs the particle's motion.  Einstein gestured
tentatively toward such a theory of light already in his 1905 paper,
and came to endorse such a picture much more openly (though never
fully in publication) in the subsequent decades.    
Eugene Wigner, for example, reports that Einstein
\begin{quote}
``was very early well aware of the wave-particle
duality of the behavior of light (and also of particles); in their
propagation they show a wave character and show, in particular,
interference effects.  Their emission and absorption are
instantaneous, they behave at these events like particles.  In order
to explain this duality of their behavior, Einstein proposed the idea
of a `guiding field' (\emph{F\"uhrungsfeld}).  This field obeys the
field equations for light, that is Maxwell's equation.  However the
field only serves to \emph{guide} the light quanta or particles, they
move into the regions where the intensity of the field is high.'' 
\cite{wigner}
\end{quote}
One of Einstein's early biographers, Philipp Frank, similarly reports
that in ``conversation Einstein expressed this dual charater of
light as follows:  `Somewhere in the continuous light waves there are
certain `peas', the light quanta'.''  \cite{frank}

Given the simplicity and naturalness of this way of understanding the
empirical wave-particle duality, it is not surprising that many other
physicists picked up and developed -- or independently arrived at --
the same kind of picture.  Hendrik Lorentz, for example, still advocated
in 1927 (what de Broglie had dubbed) a ``pilot-wave'' model of light, 
and credited the idea to Einstein:
\begin{quote}
``Can the [wave and particle characters of
light] be reconciled?  I should like to put forward some
considerations about this question, but I must first say that Einstein
is to be given credit for whatever in them may be sound.  As I know
his ideas concerning the points to be discussed only by verbal
communication, however, and even by hearsay, I have to take the
responsibility for all that remains unsatisfactory.''  \cite{lorentz}
\end{quote}
Lorentz then goes on to develop a precise -- though ultimately
untenable -- mathematical formulation of this model.

John Slater reports that, several years earlier, he and many others
had also been working on the same ideas:  a
\begin{quote}
``number of scientists -- W.F.G. Swann
among others -- had suggested that the purpose of the electromagnetic
field was not to carry a continuously distributed density of energy,
but to guide the photons in some manner.  This was the point of view
which appealed to me, and during my period at the Cavendish Laboratory
in the fall of 1923, I elaborated it.''  \cite{slater1,slater2}
\end{quote}
Curiously, though, these early pilot-wave models of light rarely made
it into publication, and seem to have been largely forgotten.

Part of the reason for this is the emergence and ascendancy of the 
Copenhagen ``tranquilizing philosophy'' (as Einstein once called
Bohr's ideology of Complementarity). \cite{tranq} 
Slater, for example, reports
that the pilot-wave ontology found no sympathy with Bohr and
his colleagues, and was eventually just lost in the rising tide of the
Copenhagen hegemony:
\begin{quote}
``As soon as I discussed [these ideas] with Bohr and
Kramers, I found them enthusiastic about the idea of the
electromagnetic waves emitted by oscillators during the stationary
states... But to my consternation I found that they completely refused
to admit the real existence of the photons.  It had never occurred to
me that they would object to what seemed like so obvious a deduction
from many types of experiments.  ....  This conflict, in which I
acquiesced to their point of view but by no means was convinced by any
arguments they tried to bring up, led to a great coolness between me
and Bohr, which was never completely removed.''  \cite{slater2}
\end{quote}
Of course, around this same time, Louis de Broglie had proposed
extending the wave-particle duality -- by then a clear empirical fact
for light -- also to electrons and other material particles.  Einstein
remarked that de Broglie had thus ``lifted a corner of the great
veil.''  \cite[p 43]{crossroads}  As Slater tells it, de Broglie's 
\begin{quote}
``point of view about the relation of photons and the
electromagnetic field was essentially the same one to which I had come
practically simultaneously.  But he did not have the antagonism of
Bohr to contend with, and consequently he followed his ideas to their
obvious conclusion.  If there were an electromagnetic wave to guide
the scattered photon in the Compton effect, why should there not also
be a wave of some sort to guide the recoil electron?  The two were
inextricably tied together.  Thus came the origin of wave
mechanics.''  \cite{slater2}
\end{quote}
Wave mechanics reached its culmination in 1926, when Schr\"odinger 
developed the dynamical time-evolution equation for de Broglie's 
electron waves.  Although Schr\"odinger himself didn't favor a
pilot-wave ontology, but instead wanted to invest the wave function
\emph{alone} with physical reality, several physicists -- 
Einstein, de Broglie, and presumably others -- were working during
this same period to construct a full, mathematically-precise
pilot-wave theory for electrons.  Indeed, Einstein
developed such a theory, which he presented in May of 1927 at a meeting of
the Prussian Academy of Sciences, and which he (it seems) also intended to present 
at the 1927 Solvay conference.  But he retracted his paper (which was 
then never published) at the last minute.  \cite{belousek, holland}
(See also section 11.3 of Ref. \cite{crossroads}.)
De Broglie's pilot-wave theory for electrons \emph{was} published and 
subsequently presented 
at Solvay in 1927, and was the subject of extensive discussion 
there.  \cite{crossroads}

As is well-known, however, shortly after 1927, the pilot-wave ontology
(now for electrons) also sank into obscurity.  This is in no small part due,
again, to the rising influence of Bohr and the Copenhagen approach to
quantum theory. \cite{cushing}  De Broglie himself rather tragically 
abandoned his own theory,  based on some combination of failing to 
understand fully the implications of his own ideas, and
what seems to have amounted to intense peer
pressure.  David Bohm, 25 years later in 1952, independently
rediscovered and further developed the pilot-wave approach to
(non-relativistic, many particle) quantum mechanics.  \cite{bohm}
But Bohm's theory too remained outside the scientific mainstream.  

The de Broglie - Bohm theory (a.k.a. ``Bohmian Mechanics'') 
has enjoyed, in recent decades, renewed
attention, triggered largely by the support of J.S. Bell and in
particular by the role played by the pilot-wave theory in stimulating 
the thinking that led to Bell's Theorem.  But the theory is still not
widely understood, taught, or appreciated by physicists.  
The philosophical, historical, and cultural reasons for this rather
curious state of affairs (curious, that is, given the naturalness of
the pilot-wave approach to understanding the empirical wave-particle
duality) has been explored in Ref. \cite{cushing}.  

Of primary interest to us here, however, is a certain more technical
issue which seems to have played an important role in, at least,
Einstein's assessment of his own and de Broglie's (and later, Bohm's)
versions of the pilot-wave theory -- and which, indeed, continues to
feature prominently in polemics against the pilot-wave ontology.  

The issue is this:  those (like Slater) who were initially
sympathetic to the pilot-wave ontology no
doubt expected that, for a system of $N$ particles moving in three
spatial dimensions, the theoretical description would be of $N$
wave-particle pairs -- each pair consisting of a point particle guided
in some way by an associated wave propagating in 3-space.  (Presumably
there would also exist, in the general case, dynamical interactions among
the wave-particle pairs.)

But Schr\"odinger's wave function for such an $N$-particle system was
emphatically \emph{not} a set of $N$ (interacting) waves, each propagating in
3-space.  It was, rather, a \emph{single} wave propagating in the
$3N$-dimensional \emph{configuration space} for the system.  De
Broglie's 1927 pilot-wave theory simply inherited this feature:  it
was a theory of particles being guided through physical 3-space by a
wave, yes, but a very strange and seemingly too-abstract wave which
didn't live in 3-space at all.  

J.S. Bell introduced the term ``beables'' (a deliberate contrast to
the vaguely-defined ``observables'' which, he thought, played too 
prominent a role in orthodox, Copenhagen quantum theory) to name
whatever is posited, by a candidate theory, as corresponding directly
to something that is physically real (independent of any
``observation'').  \cite[p 52]{bell}
He then divides beables into two categories, local and non-local:
\begin{quote}
``\emph{Local} beables are those which are definitely associated with
particular space-time regions.  The electric and magnetic fields of
classical electromagnetism, ${\bf{E}}(t,x)$ and ${\bf{B}}(t,x)$ are again
examples, and so are integrals of them over limited space-time
regions.  The total energy in all space, on the other hand, may be a
beable, but is certainly not a local one.''   \cite[p 234-5]{bell}
\end{quote}
Actually, Bell's example here leaves something to be desired.  In
classical electromagnetism, it is dubious to regard
integrals of the fields (or even, for example, the electromagnetic
energy density at a point) as beables.  Such quantities certainly in
some sense \emph{exist} (according to the theory), but they aren't, in
standard readings of the theory at least, supposed to directly describe
physical reality in the same way that the fields themselves do.  They
are the sorts of things theorists may be interested in
\emph{calculating}, but not the sorts of things that physical reality
itself is (according to the theory) \emph{made of}.  
Bell's example of a non-local beable -- ``the total energy in all
space'' -- also doesn't seem quite right.  This is certainly a non-local
quantity, in the sense he explains, but it doesn't seem plausible to
grant it ``beable status'' \cite[p 53]{bell} for the theory in question.

Probably the reason for the confusing 
examples is that Bell is trying to do the impossible:  to
give a familiar example of a non-local beable from an intuitively
clear, classical theory.  But, arguably, no such objects exist in any such
theories.  

In any case, the utility of the local vs. non-local beable distinction
becomes clear in the context of quantum theories which include, among
the beables, a wave function which lives, not in 3-space, but
\begin{quote}
``in a much bigger space, of $3N$-dimensions.  It makes no sense
to ask for the amplitude or phase or whatever of the wavefunction at a
point in ordinary space.  It has neither amplitude nor phase nor
anything else until a multitude of points in ordinary three-space are
specified.''  \cite[p 204]{bell}
\end{quote}
For any theory in which the wave function has beable status, then, it
is necessarily a non-local beable.  And this
provides a convenient alternative way to state what is surprising and
unfamiliar about the de Broglie - Bohm pilot-wave theory:  in addition to
positing local beables (the particles), the theory also posits a
genuinely non-local beable (the configuration space wave function 
which pilots them). 

Don Howard has argued that \emph{this} (and/or the
intimately related issue Howard dubs ``non-separability'') was Einstein's
primary concern with his own and de Broglie's 1927 pilot-wave
theories.  Indeed, the non-local beable character of Schr\"odinger's 
wave function stood out to Einstein from the very beginning.  Here he 
is, in a letter to Ehrenfest of April 12, 1926, praising
Schr\"odinger's wave
mechanics in comparison to Heisenberg's matrix mechanics, but also
noting a curious feature of the former:
\begin{quote}
``The Born-Heisenberg thing will certainly not be right.  It appears
not to be possible to arrange uniquely the correspondence of a matrix
function to an ordinary one.  Nevertheless, a mechanical problem is
supposed to correspond uniquely to a matrix problem.  On the other
hand, Schr\"odinger has constructed a highly ingenious theory of
quantum states of an entirely different kind, in which he lets the De
Broglie waves play in phase space.  The things appear in the Annalen.
No such infernal machine, but a clear idea and -- `compelling' in its
application.''  \cite[p 82-3]{howard}
\end{quote}
But, as Howard notes, ``Einstein's enthusiasm was short lived.''  On
the first of May, he writes to Lorentz:
\begin{quote}
``Schr\"odinger's conception of the quantum rules makes a great
impression on me; it seems to me to be a bit of reality, however
unclear the sense of waves in n-dimensional q-space remains.''
\cite[p 83]{howard} 
\end{quote}
And here he is on June 18, in a letter again to Ehrenfest:
\begin{quote}
``Schr\"odinger's works are wonderful -- but even so one nevertheless
hardly comes closer to a real understanding.  The field in a
many-dimensional coordinate space does not smell like something
real.''   \cite[p 83]{howard}
\end{quote}
It didn't take long for his discomfort with this curious aspect of the
de Broglie - Schr\"odinger wave to turn into a research program --
which program would ultimately be incorporated into Einstein's
infamous and fruitless search for a ``unified field theory''.  On
June 22, Einstein writes to Lorentz:
\begin{quote}
``The method of Schr\"odinger seems indeed more correctly conceived
than that of Heisenberg, and yet it is hard to place a function in
coordinate space and view it as an equivalent for a motion.  But if
one could succeed in doing something similar in four-dimensional
space, then it would be more satisfying.''  \cite[p 83]{howard}
\end{quote}
This thought is repeated in an August 21 letter to Sommerfeld:
\begin{quote}
``Of the new attempts to obtain a deeper formulation of the quantum
laws, that by Schr\"odinger pleases me most.  If only the undulatory
fields introduced there could be transplanted from the n-dimensional
coordinate space to the 3 or 4 dimensional!'' \cite{howard}
\end{quote}
And yet again in an August 28 letter to Ehrenfest:
\begin{quote}
``Schr\"odinger is, in the beginning, very captivating.  But the waves
in n-dimensional coordinate space are indigestible...''  \cite{howard}
\end{quote}
It is clear, then, that the ``indigestibility'' of a physically real
wave on an
abstract, high-dimensional space was a primary roadblock for
Einstein's acceptance of Schr\"odinger's wave mechanics.  \emph{This}
is evidently the feature he had in mind when, in a January 11, 1927 letter to
Ehrenfest, he described the ``Schr\"odinger business'' as ``noncausal
and altogether too primitive'' and when, on February 16 in a letter to
Lorentz, he denied emphatically that the quantum theory could ``be the
description of a real process.''  \cite[p 84]{howard}  

It stands to reason, then, that this same feature of de Broglie's
mature 1927 pilot-wave theory (which simply incorporates Schr\"odinger's
configuration space wave function) stands behind Einstein's (perhaps otherwise
surprisingly) cool reaction to that theory -- and also his downright
cold reaction to Bohm's theory 25 years later.  \cite{cheap}

De Broglie himself also seems to have had reservations about this aspect of
his theory.  Valentini and Bacciagaluppi summarize him, in his (May
1927) first full published presentation of his pilot-wave theory, as 
asserting
``that configuration space is `purely abstract', and that a wave
propagating in this space cannot be a physical wave: instead, the
physical picture of the system must involve $N$ waves propagating in
3-space.''  \cite[p 68]{crossroads}
Later, at the 1927 Solvay conference, de Broglie states:   ``It
appears to us certain that if one wants to \emph{physically} represent
the evolution of a system of $N$ corpuscles, one must consider the
propagation of $N$ waves in space...''  \cite[p 79]{crossroads}
Unlike Einstein, however, de Broglie (at least at this
time) didn't seem to appreciate the non-trivial character of the
problem.  That is, de Broglie seemed to think that it was somehow
unproblematic or straightforward to interpret Schr\"odinger's
configuration-space wave as some kind of abstract description of the
desired ``$N$ waves propagating in 3-space.''  
\cite[p 69-84]{crossroads}

It is not clear what
role de Broglie's eventual realization that this problem was in fact
not trivial at all, might have played in his decision to give up his theory
shortly after 1927.  It is clear, however -- and perhaps suggestive --
that de Broglie began working on this particular issue immediately
after his interest in the pilot-wave
ontology was rekindled by Bohm's work in 1952.  See, for example,
Chapter VI of Ref. \cite{deb}.  See also Ref. \cite{freistadt} for a
helpful review of the various attempts by de Broglie, Vigier, and
others in this direction during this period.\footnote{It should 
perhaps be noted that
these attempts, as reviewed in particular in Section 5 of Ref. 
\cite{freistadt}, seem to be rather muddled and unsuccessful.  For
example, depending on precisely how one interprets the meaning of the
symbol $r$ that appears (for example) in Equation (5.5), this and
other Equations are either trivially wrong or valid (but in a way that
severely obfuscates the nature and extent of the difficulty).  A more
careful and extensive review of these historical attempts at creating
(what is here being called) a TELB -- and their relation to the scheme
proposed in the current paper -- would I think be worthwhile.}

There is reason to think that this issue was a central concern for
others (besides Einstein and de Broglie) as well.  As Linda Wessels
reports in recounting ``Schr\"odinger's Route to Wave Mechanics''
\begin{quote}
``In the
case of a single classical particle $\psi$ could be interpreted as a
wave function describing a matter wave.  For a system of $n$ classical
particles, however, $\psi$ was a function of $3n$ spatial coordinates
and therefore described a wave in a $3n$-dimensional space that could
not be identified with ordinary physical space.  To give his theory a
wave interpretation Schr\"odinger would either have to show how the
$\psi$ in $3n$-dimensional space determined $n$ waves in 3-dimensional
space, or reformulate the theory so that it would yield directly the
required $n$ wave functions.  (Eventually each of these escape routes
were to be explored, but neither would prove successful.)''
\cite[p 333]{wessels}
\end{quote}
Wessels then adds in a footnote, citing a 1962 interview with Carl
Eckart conducted by John Heilbron:
``The obvious solution would be to rewrite the equations of
wave mechanics so that even for a system of several `particles', only
three-dimensional wave functions would be determined.  C. Eckart has
reported that at one time he attempted this and remarked that \emph{it was
something that initially `everybody' was trying to do}.'' (Emphasis added.)

And this concern over the ``non-local beable'' character of the pilot
wave in pilot-wave theory remains, to this day, a vulnerable point for
polemics against the theory and/or the associated ontology.  For
example, two recent commentators have noted that
\begin{quote}
``If only spacetime is real, one would have to figure out a way
to write the wave function as a function of 3-space instead of
3n-space.  This would implement de Broglie's original interpretation,
in which the $\Psi$-field is conceived of as propagating in physical
space.  Although there have been some attempts at doing this, ...
none have been completely successful.  In our opinion,  if
it can be achieved, this is the most desirable option, although a
certain amount of pessimism concerning its chances is probably in
order.''  \cite{cw}
\end{quote}
And even more recently, this point was raised by N. David Mermin in
a passing dig at the theory:  advocates of the pilot-wave theory, he
suggests, must implausibly give the ``$3N$-dimensional configuration
space .... just as much physical reality as the rest of us ascribe to
ordinary three-dimensional space.''  \cite{mermin}

Let us summarize the historical background context we have been
surveying and then, finally, state the thesis of the present article.
The two central background claims here are that (i) the pilot-wave
picture is a natural and intuitively-appealing way to try
to understand the kinds of phenomena whose explanation has been, from 
the very beginning, the whole purpose of quantum theory; but (ii) the
pilot-wave theory (in which one evidently must take the wave function 
as physically real, as a beable \cite[p 128]{bell} ) has been \emph{hurt} -- that is, rejected and marginalized -- 
by the fact that, at least in extant formulations, the wave which does 
the piloting is a wave not on physical 3-space, but on an abstract
$3N$-dimensional configuration space, it being (at best)
counter-intuitive to take such an object as physically real.  

The point of the present article is to show that, after
all, it \emph{is} possible to formulate a pilot-wave theory of the
sort anticipated and then unsuccessfully sought after by Einstein, de 
Broglie, and others --
that is, a theory in which $N$ particles are guided by $N$
(interacting) fields in 3-space.  Actually, the theory to be proposed
here is not \emph{precisely} of this sort, because (as we will
explain) additional fields (also in 3-space) are involved as well --
fields which do not influence the particles directly, but only
indirectly, by exerting direct influence on the pilot-waves.  
These new fields, as it turns out, can be understood as necessary
precisely to fix the kinds of problems encountered by Einstein,
Bohr-Kramers-Slater, and other early attempts to construct a quantum 
theory without non-local beables:  such theories lacked what is now 
called \emph{entanglement} and so failed to predict, for example,
strict energy and momentum conservation in scattering processes. 
(See Chapter 9 of Ref. \cite{crossroads} for a fuller discussion.)

One important feature of the theory to be presented is simply that
it shows (by explicit construction) how one
can in principle explain entanglement phenomena (which are in some sense the
essence of quantum theory) without positing any such non-local beable
as a wave function living on a high-dimensional space.  Unfortunately,
what one \emph{does} need to posit may seem, by comparison,
extravagant to the point of implausibility.  So be it.  The theory of
exclusively local beables (TELB) put forward here is only intended as
an un-serious toy model, to illustrate 
in principle that the kind of theory envisioned
by (among others) Einstein and de Broglie can, after all, be 
constructed.  It should thus be considered merely as a possible 
jumping-off point for those interested in picking up and developing
this long-since- (but, it seems, prematurely-) abandoned idea. 
  
Let us then turn to seeing how this trick can be done.

\section{Qualitative Overview}

We begin with the de Broglie - Bohm pilot-wave theory for (spinless)
particles.  This is a theory which, as mentioned, contains both local
and non-local beables -- the particles and pilot-wave, respectively.

For simplicity and to make certain features more
intuitively graspable, we consider the theory of $N=2$ particles which
move in a single spatial dimension.  Then the configuration space for
the system is two-dimensional.  The time-evolution of the wave
function is given, as usual, by Schr\"odinger's equation:
\begin{equation}
i \hbar \frac{\partial \Psi(x_1,x_2,t)}{\partial t} = -
\frac{\hbar^2}{2m_1} \frac{\partial^2 \Psi(x_1,x_2,t)}{\partial x_1^2}
- 
\frac{\hbar^2}{2m_2} \frac{\partial^2 \Psi(x_1,x_2,t)}{\partial x_2^2}
+ V\left[x_1,x_2,t\right] \, \Psi(x_1,x_2,t)
\label{scheq}
\end{equation}
where $m_1$ and $m_2$ are, respectively, the masses of particles $1$
and $2$ and $V$ is the potential energy associated with the configuration
$(x_1,x_2)$ at time $t$.  We also assume, for reasons that will become
obvious, that the wave function is analytic everywhere.

The time-evolution of the particle positions $X_1(t)$ and $X_2(t)$ 
is determined by the gradient, along the appropriate direction in
configuration space, of the phase of $\Psi$ at the actual
configuration point:
\begin{equation}
\frac{d X_i(t)}{dt} = \frac{\hbar}{m_i} \, \text{Im} \left( \frac{\nabla_i
    \Psi}{\Psi} \right) \Big|_{x_1 = X_1(t), \, x_2 = X_2(t)}.
\end{equation}
The existence, in the theory, of the ``actual configuration'' --
represented by $X_1(t)$ and $X_2(t)$ -- allows one to define the
so-called ``conditional wave-function'' for each particle.  \cite{au}
This is,
for a given particle, simply the full, configuration-space wave
function evaluated at the actual location of the other particle.
Thus,
\begin{equation}
\psi_1(x,t) \equiv \Psi(x,x_2,t)\big|_{x_2 = X_2(t)}
\label{cwf1}
\end{equation}
and
\begin{equation}
\psi_2(x,t) \equiv \Psi(x_1,x,t)\big|_{x_1 = X_1(t)}.
\label{cwf2}
\end{equation}
The idea of using such conditional wave functions as part of the
ontology of a pilot-wave theory is not new.  \cite[p
79-80]{crossroads}  This option is usually abandoned,
however, based on an argument that an ontology consisting
\emph{exclusively} of the particle positions and the conditional wave
functions, cannot work.  (See, for example, Ref. \cite{wigner}.)
Let us review this.

The time-evolution law for each particle position can be written in
terms of the associated conditional wave function as follows:
\begin{equation}
\frac{dX_i(t)}{dt} = \frac{\hbar}{m_i} \, \text{Im} \left( \frac{\partial
    \psi_i(x,t) / \partial x}{\psi_i(x,t)} \right) \Big|_{x = X_i(t)}.
\label{particlevelocity}
\end{equation}
But the time-evolution of the conditional wave functions themselves
depends not only on the structure of those conditional wave functions
(and the particle positions), but also on the full,
configuration-space wave function from which the conditional wave
functions were extracted.  \cite{densitymatrix}  

We illustrate this here with a simple example that will be useful also
in later discussions.  Consider a simple scattering experiment (in our
impoverished, 2-dimensional configuration space):  particle $1$,
following a localized wave-packet, is projected toward particle $2$,
which is at rest (in a stationary wave packet) near the origin.  The
two particles have some kind of contact interaction, e.g.,
$V(x_1,x_2,t) \sim \delta(x_1 - x_2)$.  Assume the strength of the
interaction is chosen (relative to the energies and other properties
of the two particles) so that the two possible outcomes -- particle
$1$ ``tunnels'' past particle $2$ and continues on its way, or
particle $1$ collides with particle $2$ and stops while particle $2$
moves off to the right -- are (according to
ordinary QM) equally probable.

\begin{figure}[t]
\centering
\scalebox{1.0}{
% Generated with LaTeXDraw 1.9.5
% Fri Aug 28 13:15:43 EDT 2009
% \usepackage[usenames,dvipsnames]{pstricks}
% \usepackage{epsfig}
% \usepackage{pst-grad} % For gradients
% \usepackage{pst-plot} % For axes
\scalebox{1} % Change this value to rescale the drawing.
{
\begin{pspicture}(0,-4.1492186)(8.501875,4.1692185)
\definecolor{color51b}{rgb}{0.4,0.4,0.4}
\definecolor{color81}{rgb}{0.2,0.2,0.2}
\definecolor{color123b}{rgb}{0.8,0.8,0.8}
\psline[linewidth=0.04cm,arrowsize=0.05291667cm 2.0,arrowlength=1.4,arrowinset=0.4]{<->}(4.1,4.070781)(4.1,-4.1292186)
\psline[linewidth=0.04cm,arrowsize=0.05291667cm 2.0,arrowlength=1.4,arrowinset=0.4]{<->}(0.0,-0.02921875)(8.2,-0.02921875)
\usefont{T1}{ptm}{m}{n}
\rput(8.021406,-0.41921875){$x_1$}
\usefont{T1}{ptm}{m}{n}
\rput(3.6214063,3.9807813){$x_2$}
\pscircle[linewidth=0.01,dimen=outer,fillstyle=solid,fillcolor=color51b](1.4,-0.02921875){0.7}
\psdots[dotsize=0.16](1.3,0.47078124)
\psline[linewidth=0.02cm,linecolor=color81,linestyle=dashed,dash=0.16cm 0.16cm](1.3,3.5707812)(1.3,-3.8292189)
\psline[linewidth=0.04cm,linestyle=dotted,dotsep=0.16cm](0.3,-3.8292189)(7.9,3.7707813)
\psline[linewidth=0.06cm,arrowsize=0.05291667cm 3.62,arrowlength=1.38,arrowinset=0.4]{->}(1.2,-0.02921875)(1.8,-0.02921875)
\pscircle[linewidth=0.01,dimen=outer,fillstyle=solid,fillcolor=color123b](6.8,-0.02921875){0.7}
\pscircle[linewidth=0.01,dimen=outer,fillstyle=solid,fillcolor=color123b](4.1,2.6707811){0.7}
\psline[linewidth=0.06cm,arrowsize=0.05291667cm 3.62,arrowlength=1.38,arrowinset=0.4]{->}(6.6,-0.02921875)(7.2,-0.02921875)
\psline[linewidth=0.06cm,arrowsize=0.05291667cm 3.62,arrowlength=1.38,arrowinset=0.4]{->}(4.1,2.4707813)(4.1,3.0707812)
\psdots[dotsize=0.16](4.4,2.4707813)
\psline[linewidth=0.084cm,linecolor=color123b](7.5,3.3707812)(0.7,-3.4292188)
\psline[linewidth=0.02cm,linecolor=color51b,linestyle=dashed,dash=0.16cm 0.16cm](4.4,3.6707811)(4.4,-3.9292188)
\psline[linewidth=0.02cm,linecolor=color51b,linestyle=dashed,dash=0.16cm 0.16cm](0.3,2.4707813)(8.0,2.4707813)
\psline[linewidth=0.02cm,linecolor=color81,linestyle=dashed,dash=0.16cm 0.16cm](0.2,0.47078124)(7.9,0.47078124)
\end{pspicture} 
}
}
\caption{
A scattering experiment involving two interacting particles in one 
spatial dimension:  particle 1 is projected toward particle 2, which is 
initially at rest at the origin.  A contact potential between the two
particles (represented by the diagonal grey line in the figure) causes the
initial (dark grey) wave packet in configuration space to split into
non-overlapping final packets (represented by the two light grey
blobs).  The actual configuration point at the beginning of the
experiment is represented by the black dot in the dark grey blob; 
the initial conditional wave functions for the two particles are
simply the full wave function evaluated along (respectively) the two dashed
lines through the actual configuration point.  
(And similarly for the final conditional wave functions.)
We suppose that the 
initial conditions are such that this initial configuration
leads, by the deterministic pilot-wave dynamics, to the final
configuration represented by the other black dot.  This
corresponds, in 1-D physical space, to particle $1$ having suffered a
billiard-ball like collision in which it stops completely and its
momentum is transferred to particle $2$, which moves off to the
right (in physical space, not configuration space!).  
}
\label{fig1}
\end{figure}
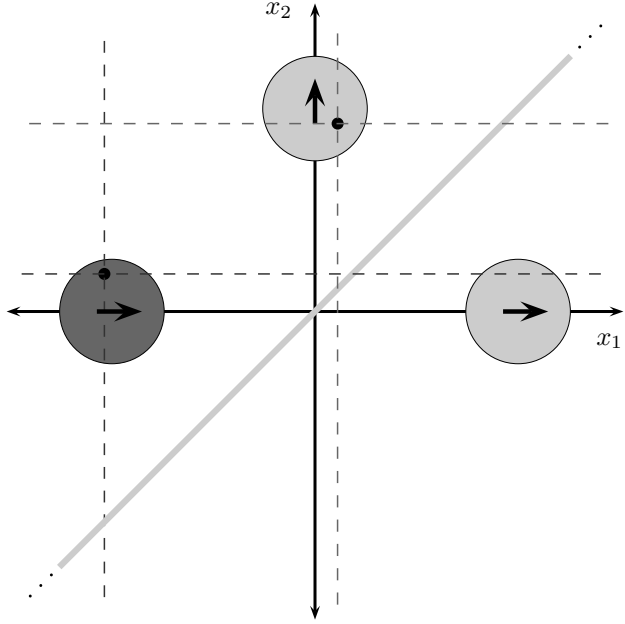

The dynamical sequence, in the 2-dimensional
configuration space, is illustrated in Figure 1.  Here we suppose that
the initial particle positions are such as to produce the second
possible outcome:  there is a billiard-ball-like collision in which
particle $1$ is brought to rest and its momentum is transfered to
particle $2$, which moves off.

Now the point is that (without changing either the initial particle
positions \emph{or the initial conditional wave functions}) the
initial  state of the full, configuration-space wave function could 
have been changed so as to produce a dramatically different outcome.
For example, had the initial state for $\Psi$ involved an
appropriate superposition
of incoming waves in the configuration space -- as illustrated in
Figure 2 -- destructive interference would absolutely prohibit the
outcome considered in the first experiment, and would instead ensure
that the actual configuration point emerges to the right in the
figure, i.e., that particle $1$ simply tunnels through particle $2$
and continues its initial motion to the right.

Comparison of the two examples thus suggests that the particle positions
and conditional wave functions \emph{alone} cannot be a sufficient
ontology for a theory -- different things can happen (according to the
full, configuration-space dynamics) even though the initial particle
positions and conditional wave functions are (in the two examples)
identical.  The comparison also brings out precisely what aspect is
missing from the proposed ontology:  the conditional wave functions
necessarily fail to capture information about the structure of the
(full, configuration space) wave function from regions that are
``diagonal'' from the actual configuration point.  Such information
will exist whenever the wave function fails to factorize, i.e.,
whenever the quantum state is entangled.  And of course -- as the
above comparison illustrates -- such information can be dynamically
relevant to the motion of the particles (and indeed also the evolution
of the conditional wave functions).  So any proposed new ontology has
to find a way to capture this information if it is going to reproduce
the empirically correct predictions of the configuration space
pilot-wave theory for the particle trajectories.

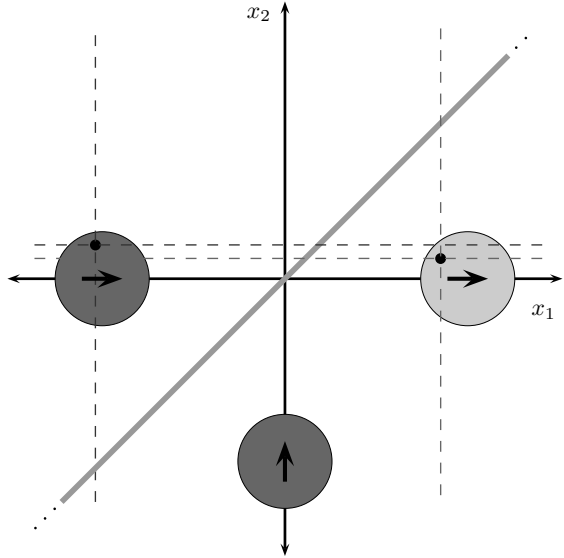
\begin{figure}[t]
\centering
\scalebox{0.9}{
% Generated with LaTeXDraw 1.9.5
% Fri Aug 28 13:37:35 EDT 2009
% \usepackage[usenames,dvipsnames]{pstricks}
% \usepackage{epsfig}
% \usepackage{pst-grad} % For gradients
% \usepackage{pst-plot} % For axes
\scalebox{1} % Change this value to rescale the drawing.
{
\begin{pspicture}(0,-4.12)(8.401875,4.12)
\definecolor{color51b}{rgb}{0.4,0.4,0.4}
\definecolor{color81}{rgb}{0.2,0.2,0.2}
\definecolor{color123b}{rgb}{0.8,0.8,0.8}
\definecolor{color111}{rgb}{0.6,0.6,0.6}
\psline[linewidth=0.04cm,arrowsize=0.05291667cm 2.0,arrowlength=1.4,arrowinset=0.4]{<->}(4.1,4.1)(4.1,-4.1)
\psline[linewidth=0.04cm,arrowsize=0.05291667cm 2.0,arrowlength=1.4,arrowinset=0.4]{<->}(0.0,0.0)(8.2,0.0)
\usefont{T1}{ptm}{m}{n}
\rput(7.9214063,-0.49){$x_1$}
\usefont{T1}{ptm}{m}{n}
\rput(3.7214062,3.91){$x_2$}
\pscircle[linewidth=0.01,dimen=outer,fillstyle=solid,fillcolor=color51b](1.4,0.0){0.7}
\psdots[dotsize=0.16](1.3,0.5)
\psline[linewidth=0.02cm,linecolor=color81,linestyle=dashed,dash=0.16cm 0.16cm](1.3,3.6)(1.3,-3.3)
\psline[linewidth=0.04cm,linestyle=dotted,dotsep=0.16cm](0.4,-3.7)(7.8,3.7)
\psline[linewidth=0.06cm,arrowsize=0.05291667cm 3.62,arrowlength=1.38,arrowinset=0.4]{->}(1.1,0.0)(1.7,0.0)
\pscircle[linewidth=0.01,dimen=outer,fillstyle=solid,fillcolor=color123b](6.8,0.0){0.7}
\psline[linewidth=0.06cm,arrowsize=0.05291667cm 3.62,arrowlength=1.38,arrowinset=0.4]{->}(6.5,0.0)(7.1,0.0)
\pscircle[linewidth=0.01,dimen=outer,fillstyle=solid,fillcolor=color51b](4.1,-2.7){0.7}
\psline[linewidth=0.06cm,arrowsize=0.05291667cm 3.62,arrowlength=1.38,arrowinset=0.4]{->}(4.1,-3.0)(4.1,-2.4)
\psdots[dotsize=0.16](6.4,0.3)
\psline[linewidth=0.084cm,linecolor=color111](7.4,3.3)(0.8,-3.3)
\psline[linewidth=0.02cm,linecolor=color51b,linestyle=dashed,dash=0.16cm 0.16cm](0.4,0.3)(7.9,0.3)
\psline[linewidth=0.02cm,linecolor=color51b,linestyle=dashed,dash=0.16cm 0.16cm](6.4,3.7)(6.4,-3.2)
\psline[linewidth=0.02cm,linecolor=color81,linestyle=dashed,dash=0.16cm 0.16cm](0.4,0.5)(7.9,0.5)
\end{pspicture} 
}
}
\caption{
A similar scattering experiment:  the situation is just as before, but
now the initial wave function represents an \emph{entangled} state for
the two particles, leading to a different outcome even though the
initial particle positions \emph{and the initial conditional wave
functions} are the same as before.   This illustrates the way in which
the conditional wave functions fail to capture information about the
structure of the (configuration space) wave function which is in a
``diagonal'' direction (in the configuration space) from the actual
configuration point -- i.e., the way in which the conditional wave
functions fail to capture \emph{entanglement}.  
}
\label{fig2}
\end{figure}

Before turning to the approach to be advocated, we consider briefly --
for contrast and historical interest -- another way one might consider
trying to capture the full relevant structure of the wave function using
fields on physical space:  instead of using the conditional wave
functions for each particle, one might form, for each particle, an
associated wave on physical space by (squaring and) projecting.  For
example, one could define a field for particle $1$ this way:
\begin{equation}
\eta_1(x,t) \equiv \int | \Psi(x,x_2,t)|^2 \, dx_2
\end{equation}
and similarly for particle $2$.  This would give, at the beginning of
the experiment shown in Figure 1, a wave packet for particle $1$
moving toward a stationary wave packet for particle $2$.  At the end
of the experiment, however, each particle's $\eta$-field would comprise
two packets -- one moving and one stationary -- corresponding to the
two possible final states of motion for each particle.  

One might reasonably attempt to 
interpret the two packets (for a given particle) as indicating 
two possible outcomes \emph{only
one of which is realized} (with the fact about which one is realized
being random and -- in particular -- independently random for each
particle). This would seem to provide a sensible story about what 
happens during the scattering process, but note that we would then
lose strict energy and momentum conservation:  it might turn out, for
example, that both particles end up at rest near the origin after the
experiment.  Theories of this sort were considered by (at least)
Einstein and Bohr, Kramers, and Slater, but ultimately given up when
it was realized that strict energy/momentum conservation was required
by experiment.  See Chapter 9 of Ref. \cite{crossroads} for a more
extended discussion.  (Note also the intimate connection between the
ontology just proposed and Schr\"odinger's own early views, as
reviewed and clarified recently in Ref. \cite{Sm}.)

With that (historically important) alternative approach out of the
way, let us return to the approach to be advocated here:  using the
conditional wave functions as the basis for an ontology of exclusively
local beables, but \emph{supplementing} the conditional wave functions
with some new (local) beables in order to capture the
dynamically-relevant ``diagonal'' structure of the full, configuration
space wave function.  

A possible way of doing this reveals itself as soon as we attempt to
extract a time-evolution law for the conditional wave functions from
the (configuration space) Schr\"odinger equation.  Let us illustrate
this using the conditional wave function for particle 1 from above.
Using Equation \eqref{cwf1}, the time-derivative of
$\psi_1(x,t)$ can be computed as follows:
\begin{equation}
\frac{\partial \psi_1(x,t)}{\partial t} = \frac{ \partial
  \Psi(x,x_2,t)}{\partial t} \Big|_{x_2 = X_2(t)}  \; + \; \frac{d
  X_2(t)}{dt} \, \frac{\partial \Psi(x,x_2,t)}{\partial x_2}
\Big|_{x_2 = X_2(t)}.
\end{equation}
The derivative in the first term is given by Equation \eqref{scheq},
two of the terms on the right hand side of which can be written in
terms of $\psi_1(x,t)$ once evaluated at $x_2 = X_2(t)$.  
Multiplying through by $i \hbar$, we end up with the
following Schr\"odinger-like equation for the time-evolution of
$\psi_1(x,t)$:
\begin{equation}
i \hbar \frac{\partial \psi_1(x,t)}{\partial t} =
- \frac{\hbar^2}{2m_1} \frac{ \partial^2 \psi_1(x,t)}{\partial x^2} +
V\left[x,X_2(t),t\right] \, \psi_1(x,t) \, +\,  A \, + \, B
\end{equation}
where
\begin{equation}
A = - \frac{\hbar^2}{2m_2} \frac{\partial^2 \Psi(x,x_2,t)}{\partial
  x_2^2} \Big|_{x_2 = X_2(t) }
\end{equation}
and
\begin{equation}
B = i \hbar \, \frac{d X_2(t)}{dt} \; \frac{\partial
  \Psi(x,x_2,t)}{\partial x_2} \Big|_{x_2 = X_2(t)}
.
\end{equation}
The point, of course, is that the terms we have called $A$ and $B$
cannot be written in terms of the local beables introduced so far (the
particle positions and the conditional wave functions).  

This could be  thought of as an argument that one cannot 
construct a TELB using the conditional wave functions.  
But in fact what the argument shows us is
precisely what additional local beables we need to \emph{add} to the 
ontology in order to have a well-defined, closed dynamics.  
Specifically, we introduce the new fields
\begin{equation}
\psi^{\,\prime}_1(x,t) 
\equiv \frac{\partial \Psi(x,x_2,t)}{\partial x_2} \Big|_{x_2 = X_2(t)}
\end{equation}
and
\begin{equation}
\psi^{\, \prime \prime}_1(x,t) 
\equiv \frac{\partial^2 \Psi(x,x_2,t)}{\partial x_2^2}
\Big|_{x_2 = X_2(t)}.
\end{equation}
Don't be fooled by the notation:  these are meant to be taken as
genuinely new fields, which certainly cannot be computed or defined in
terms of the conditional wave function $\psi_1(x,t)$.  The point of
introducing them is of course to make the time-evolution law for
$\psi_1(x,t)$ well-defined in terms of posited local beables.  It now
reads:
\begin{eqnarray}
i \hbar \frac{\partial \psi_1(x,t)}{\partial t} &=&
- \frac{\hbar^2}{2m_1} \frac{ \partial^2 \psi_1(x,t)}{\partial x^2} +
V\left[x,X_2(t),t\right] \, \psi_1(x,t) \nonumber \\
&& \; - \frac{\hbar^2}{2m_2} \psi^{\, \prime \prime}_1(x,t) \; + \;  
i \hbar \frac{d X_2(t)}{dt} \psi^{\, \prime}_1(x,t). 
\end{eqnarray}
(One should understand, here and occasionally in what follows, factors
like the $dX_2/dt$ on the right hand side as
merely a shorthand for the right hand side of Equation
\eqref{particlevelocity} with the appropriate value of $i$.)  

Of course, having introduced these new local beable fields, e.g.,
$\psi^{\, \prime}_1(x,t)$, we must now define \emph{their}
time-evolution.   But that can
be done straightforwardly (if tediously), by simply taking the
time-derivative of the relevant definition, and using again
Schr\"odinger's equation.  One can see that, for example, the
Schr\"odinger-like equation satisfied by $\psi^{\, \prime}_1(x,t)$ 
will involve, on the right-hand-side, terms including 
\begin{equation}
\frac{\partial ^3 \Psi(x,x_2,t)}{\partial x_2^3} |_{x_2 = X_2(t)}
\end{equation}
which we incorporate as a new local beable field, $\psi^{\, \prime
  \prime \prime}(x,t)$.  And
so on.  

And of course all of this is happening in parallel also for particle
$2$ and its associated fields.  In the end, we have a (countable)
infinity of fields associated with each particle:  a conditional wave
function which (directly) determines the particle's velocity at each
instant, and then a hierarchy of additional fields which influence the
conditional wave function (and one another) and can be thought of as
simply a way to capture or reproduce -- by what amounts to simple
Taylor expansion -- the information about the structure of the full,
configuration space wave function that the conditional wave functions
alone failed to capture.\footnote{Here we see clearly the importance
  of having assumed a (configuration space) wave function that is --
  for all times -- everywhere analytic.  For a smooth but non-analytic
  wave function, the Taylor expansions used here will encode a 
  different ``entanglement structure'' -- and hence imply, eventually,
  different particle trajectories -- than the usual formulation of the
  pilot-wave theory.  Thus, the scheme employed here will only
  reproduce the predictions of the usual pilot-wave theory for the
  special case of analytic wave functions.  The extent to which this
  is a serious shortcoming of the current proposal is perhaps
  debatable.  For example, it would be difficult to cite any empirical
  evidence that the usual formulation of pilot-wave theory (as opposed
  to the formulation developed here) makes
  correct predictions for situations involving smooth but non-analytic
  wave functions.  And anyway, there will turn out to be several
  unrelated -- and almost certainly \emph{more} serious --
  shortcomings, which render the precise assessment of the seriousness
  of this particular one rather moot.}

We will reflect later on several interesting features of this theory
(and/or the slightly modified version to be considered in Section
\ref{sec4}).  For now we note just one important feature:  despite
being a theory of (i.e., formulated in terms of) exclusively local
beables, the theory is manifestly non-local.  For example, the rate of
change of $\psi_1$ (at every point in space!) is influenced by the
instantaneous velocity of particle $2$ (or, perhaps more accurately, by
the imaginary part of the gradient of the logarithm of particle $2$'s
pilot-wave field at the location of particle $2$).  Not surprisingly, similar
dynamical non-localities will appear in the equations defining the
time-evolution of the other fields.  This non-local causation is of
course essential to the theory's ability to reproduce the predictions
of ordinary pilot-wave theory, i.e., essential to its ability to be
(in light of Bell's theorem and the associated experiments)
empirically viable.  
\footnote{Shelly Goldstein (private communication) points out that, in
  the presence of such dynamical non-localities, the distinction
  between local and non-local beables becomes rather fuzzy:  one
  could, for example, take ordinary Bohmian Mechanics and just say
  that the universal wave function lives at some particular point in
  3-space.  The wave function's choreography of the particle
  trajectories would then involve dynamical non-locality, but the
  theory would be a TELB.  Somehow, though, this strikes one as
  cheating.  Probably the underlying intuitions relate to the
  suggestion, from the very end of this paper, that we already have --
  from pre-theoretical interpretations of certain key experiments --
  some qualitative sense of what at least some of
  the local beables ought to be.}

This is admittedly a complicated, ugly, and highly contrived theory.
(And although it is straightforward to generalize from 2 particles 
moving in 1 spatial dimension to $N$ particles moving in 3 spatial 
dimensions, the complexity and ugliness in that more serious
context is surely much worse!) 
But it nevertheless represents a way of doing what many physicists
took to be important but impossible, so we believe it is worth taking
somewhat seriously -- at least to the point of asking how one
could do better.  In the following section, we will 
briefly summarize in a
more formal way the theory's ontology and its defining equations.  A
later section will then present some ideas for how to use the same
basic qualitative approach to construct a (marginally) more plausible
theory.

\section{A complicated, ugly, and highly contrived TELB}
\label{sec3}

The previous section explained how a theory of exclusively local
beables (TELB) can be constructed, ``top-down'', by starting with the
ordinary pilot-wave theory (with its non-local beable, a wave function
on configuration space).  Here, we take the opposite approach to
presenting the theory and simply lay out the theory's ontology and
dynamics.  What makes this interesting is, of course, that the theory
to be presented simply does not include anything like the ordinary
quantum mechanical wave function (on configuration space) but is
instead formulated exclusively in terms of local beables.  Yet, as the
previous section should have made clear, the theory is (by
construction) empirically equivalent to the de Broglie - Bohm pilot
wave theory (at least, as noted, for the case of analytic wave
functions) and hence also to ordinary quantum theory.
\footnote{Actually, the question of empirical equivalency presupposes
  also some constraints on allowed initial conditions -- an issue we
  intend to gloss over here.  We will discuss this important issue
  in more detail in the following section, where we present a distinct
  but related theory which might be taken at least a little
  more seriously.}

The ontology of the theory (again, for two particles moving in one
spatial dimension) is as follows:  each ``particle'' (in the informal
way of speaking) comprises a literal particle (a point moving with
some definite trajectory through physical space), an associated
pilot-wave field, and an infinite set of what might be termed ``entanglement
fields''.   (For the more general case of $N$ particles, the taxonomy
of ``entanglement fields'' is a little more complicated, but still
basically straightforward.)   It is assumed that the image, in the
theory, of the familiar material world is to be found in the particles
-- that is, somehow, we can \emph{see} the particles, but the
pilot-wave and entanglement fields are (like the electric and magnetic
fields of classical E\&M) \emph{invisible}.  

The particles move according to the following law:
\begin{equation}
\frac{dX_i(t)}{dt} = \frac{\hbar}{m_i} \text{Im} \left(
  \frac{\partial \psi_i(x,t) / \partial x}{\psi_i(x,t)}\right) \Big|_{x = X_i(t)}
\end{equation}
while the pilot-wave field for particle 1 evolves in 
time according to:
\begin{eqnarray}
i \hbar \frac{\partial \psi_1(x,t)}{\partial t} &=&
- \frac{\hbar^2}{2m_1} \frac{ \partial^2 \psi_1(x,t)}{\partial x^2} +
V\left[x,X_2(t),t\right] \, \psi_1(x,t) \nonumber \\
&& \; - \frac{\hbar^2}{2m_2} \psi^{\, \prime \prime}_1(x,t) \; + \;  
i \hbar \frac{d X_2(t)}{dt} \psi^{\, \prime}_1(x,t). 
\end{eqnarray}
Finally, the particle 1 entanglement fields $\psi^{\, \prime}_1(x,t)$, 
$\psi^{\, \prime \prime}_i(x,t)$, $\psi^{\, \prime \prime \prime}_i(x,t)$, 
etc., evolve according to:
\begin{eqnarray}
i \hbar \frac{ \partial \psi^{(n)}_1(x,t)}{\partial t} &=& 
-\frac{\hbar^2}{2m_1} \frac{\partial^2 \psi^{(n)}_1(x,t)}{\partial
  x^2} - \frac{\hbar^2}{2m_2} \psi^{(n+2)}_1(x,t) \nonumber \\
&& \; + i \hbar \, \frac{dX_2(t)}{dt} \psi^{(n+1)}_1(x,t) \; + \; P_n
\end{eqnarray}
where the potential term $P$ is
\begin{equation}
P_n  \equiv  \sum_{i=0}^n \binom{n}{i} \, \frac{\partial^i V}{\partial
  x_2^i} \left[ x,X_2(t),t \right] \psi^{(n-i)}_1(x,t).
\end{equation}
And of course one has analogous time-evolution equations for the
fields associated with particle 2.  

This system of equations evidently describes a well-defined, closed
dynamical system in the sense that providing 
initial conditions for all of the
posited beables determines a unique state for all the beables at all
future times.  With initial conditions (especially for the
``entanglement fields'') carefully chosen to reproduce
the quantum mechanical structure -- and in particular with
appropriately random initial positions for the particles -- the 
system will evidently reproduce exactly the statistical predictions of
ordinary quantum theory (at least for analytic configuration-space
wave functions).  

It is of course obvious that we have generated this theory by deducing
it, in the manner described in the previous section, from ordinary
pilot-wave theory.  And as long as one keeps that origin in mind, it
may seem merely like a (pointless and cumbersome) mathematical
reformulation of that theory.  In order to appreciate what is
interesting about the theory, therefore, one should attempt to imagine
that it (or something like it)
could have come about (in, say, 1926 or 1927) from the
conflict between (i) the intuitively-appealing pilot-wave ontology (with 
the assumption of pilot-waves in physical space) and (ii) the growing
realization that something additional would have to be added to such a
theory in order to ensure strict energy-momentum conservation in
individual processes (and other empirically-observable features
related to what is now termed ``entanglement'').  

From \emph{this} perspective -- and to whatever extent one finds this
a plausible thing to imagine -- the proposed theory sheds an
interesting new light on ordinary quantum theory and in particular the
status of the wave function therein.  For one could imagine, after the
present theory had been proposed and tested, mathematical explorations
of its structure and predictions revealing the possibility of a
mathematically-equivalent formulation in terms of a single abstract
pilot-wave on configuration space.  From this perspective, the (in our
world, familiar) configuration space wave function would be merely a
convenient mathematical device, analogous to Hamilton's principal
function in classical mechanics -- an abstract mathematical quantity
which perhaps in some situations makes calculations simpler or more elegant,
but which one needn't take as indicating the real existence of
anything like a physically real field on configuration space.

\section{A Marginally Improved TELB}
\label{sec4}

One of the implausible features of the theory sketched in the previous
section is that an incredible fine-tuning of initial conditions 
is required to reproduce the desired quantum mechanical predictions.  
In effect, one has to remember that all of the fields have a 
certain relation to the configuration space wave function, and 
choose their initial values on this basis.  For example, 
even in a situation (like the
experiment discussed before and pictured in Figure 1) where there 
is initially no entanglement, all of the primed
``entanglement fields'' from
the theory in the previous section will be (even initially) non-zero
and will require the sort of implausibly fine-tuned initial conditions
just mentioned.

A little thought reveals, though, that when there is no entanglement
between the particles the ``entanglement fields'' defined in the previous
section are all simply proportional to the corresponding pilot-wave
fields.  For example, assuming the configuration space wave function
has a non-entangled, product structure 
\begin{equation}
\psi(x_1,x_2) = \alpha(x_1) \beta(x_2)
\end{equation}
it follows that the first-order entanglement field for particle 1 is
just proportional to $\psi_1(x) = \alpha(x) \beta(X_2(t))$:
\begin{equation}
\psi^{\, \prime}_1(x) \equiv \frac{\partial \psi(x,x_2)}{\partial x_2} 
\Big|_{x_2 = X_2(t)} = \alpha(x) \frac{\partial \beta(x_2)}{\partial x_2}
\Big|_{x_2 = X_2(t)}. 
\end{equation}
It is thus straightforward to construct alternative ``entanglement
fields'' which actually deserve the name, in the sense that they 
\emph{vanish} when there is no entanglement.  We thus introduce
\begin{equation}
\bar{\psi}_1(x,t) = \psi^{\, \prime}_1(x,t) - R_1(t) \, \psi_1(x,t)
\end{equation}
and
\begin{equation}
\bar{\bar{\psi}}_1(x,t) = \psi^{\, \prime \prime}_1(x,t) - R_2(t) \,
\psi_1(x,t) 
\end{equation}
and so on.  The proportionality factors $R$ can be thought of as 
\begin{equation}
R_n(t) \equiv \frac{ \psi^{(n)}_1(x,t)}{\psi_1(x,t)} \Big|_{x = X_1(t)}
\end{equation}
though it is actually better to take the following equivalent
statement as their definition:
\begin{equation}
R_n(t) \equiv \frac{ \partial^n \psi_2(x,t) / \partial x^n}{\psi_2(x,t)} \Big|_{x =
  X_2(t)}.
\end{equation}
Then we can construct a theory in which the conditional wave functions
play, as before, the role of the pilot-wave fields governing the
motion of associated particles, and in which information about
entanglement between particles is captured in the ``barred''
entanglement fields which, along with the pilot-wave fields and
particles, constitute the ontology for a TELB.  

It is straightforward to rewrite the Schr\"odinger-like equations for the
pilot-wave fields in terms of the ``barred'' entanglement fields.  For
example, the law defining the time-evolution of $\psi_1(x,t)$ is:
\begin{eqnarray}
i \hbar \frac{\partial \psi_1(x,t)}{\partial t} &=& 
- \frac{\hbar^2}{2 m_1} \frac{\partial^2 \psi_1(x,t)}{\partial x^2} +
V \left[ x, X_2(t), t \right] \, \psi_1(x,t) \nonumber \\
&& \, - \frac{\hbar^2}{2m_2} \bar{\bar{\psi}}_1(x,t) + i \hbar \frac{d
  X_2(t)}{dt} \bar{\psi_1}(x,t) + f(t) \psi_1(x,t)
\label{schlike}
\end{eqnarray}
where
\begin{equation}
f(t) = - \frac{\hbar^2}{2m_2} R_2(t) + i \hbar \frac{d X_2(t)}{dt} R_1(t).
\end{equation}
The dynamical equation satisfied by the pilot-wave field for particle
2 is, of course, precisely analogous.  

The term involving $f(t)$ can be understood as producing merely
a (time-dependent) overall constant factor in $\psi_1$.  This relates
to the fact that we are not using normalized wave functions.  For
example, it clearly follows from the configuration space formulation
of the theory that $\psi_1(x,t)$ evaluated at $x = X_1(t)$ will
(always) equal $\psi_2(x,t)$ evaluated at $x = X_2(t)$.  Thus, some
continuous re-adjustment of the overall phase and normalization of the
pilot-wave fields -- unfamiliar from the point of view of one-particle
Schr\"odinger equations -- is clearly necessary here.  Of course, one
could eliminate this behavior by using, instead, normalized
conditional wave functions as the pilot-wave fields.  This would
moderately simplify the Schr\"odinger-like equations satisfied by
those fields, but would in turn complicate the definition of
appropriate entanglement fields.  We set this possibility aside for
now.  

One can also straightforwardly derive (from the configuration space
theory) the time-evolution equations satisfied by the entanglement
fields.  For example, $\bar{\psi}_1(x,t)$ will obey the following
Schr\"odinger-like equation:
\begin{eqnarray}
i \hbar \frac{\partial \bar{\psi}_1(x,t)}{\partial t} &=& 
- \frac{\hbar^2}{2m_1} \frac{\partial^2 \bar{\psi}_1(x,t)}{\partial
  x^2} + V\left[x,X_2(t),t\right] \bar{\psi}_1(x,t) \nonumber \\
&& \, - \frac{\hbar^2}{2m_2} \left( \bar{\bar{\bar{\psi}}}_1(x,t) -
  R_1(t) \bar{\bar{\psi}}_1(x,t) \right) 
\, + \, 
i \hbar \frac{d X_2(t)}{dt} \left( \bar{\bar{\psi}}_1(x,t) - R_1(t)
  \bar{\psi}_1(x,t) \right) \nonumber \\
&& \, + \left( \frac{\partial V}{\partial x_2}\left[ x, X_2(t),t
  \right] - \frac{\partial V}{\partial x_2} \left[ X_1(t), X_2(t), t
  \right] \right) \, \psi_1(x,t) \nonumber \\
&& \, + \bar{f}(t) \, \psi_1(x,t)
\label{barschlike}
\end{eqnarray}
where
\begin{equation}
\bar{f}(t) = \frac{\hbar^2}{2m_1} \frac{ \partial^2 \bar{\psi}_1 /
  \partial x^2}{\psi_1} \Big|_{x = X_1(t)} \, - \, i \hbar
  \frac{dX_1(t)}{dt} \frac{\partial \bar{\psi}_1 / \partial x}{\psi_1}
  \Big|_{x = X_1(t)} .
\end{equation}
The higher-order entanglement fields for particle 1 will obey similar
equations (which are straightforward to work out), and those for
particle 2 are analogous.

Now let's think about what happens in the scattering experiment
discussed earlier (and pictured, in configuration space, in Figure 1)
according to the theory suggested here.  At the beginning of the
experiment, there is no entanglement, so all of the (``barred'') 
entanglement fields vanish identically -- that's their initial
condition.  The pilot-wave fields for the two particles will simply be
their ordinary, one-particle wave functions (with the unusual but
minor caveat that the amplitude and phase of the one-particle wave
functions should be chosen to respect the condition mentioned
earlier).   And the particles
themselves should be placed randomly, according to the usual Born rule
prescriptions:  $P\left[X_i(0) \! = \! x \right] \sim |\psi_i(x,0)|^2$.  
This suffices to define a theory of exclusively local beables (TELB)
which should, like the theory proposed in the previous section, 
exactly reproduce the predictions of ordinary pilot-wave
theory and hence also ordinary QM, at least for the case of wave
functions which are everywhere analytic.  The present re-formulation 
is an improvement over that of the previous section because it 
makes the specification of initial conditions
(at least for situations in which, initially, no entanglement is
present) quite natural.

The dynamical equations we have presented so far already provide a
sufficient basis for understanding how things develop in time as the
experiment proceeds.  To begin with, all the entanglement fields
vanish identically, and so the only non-zero terms in Equation
\eqref{schlike}  will be the familiar ``kinetic energy'' and
``potential'' terms from the usual one-particle Schr\"odinger equation.
(There is of course also the term, discussed already, which merely 
changes the overall phase and normalization of the pilot-wave field.
But this term doesn't affect the dynamically relevant structure of the
field -- i.e., the changes it effects do not influence the motion of the
particle being piloted by this field.)
It is also worth mentioning explicitly that the ``potential'' term
involves the ``conditional potential'', i.e., the classical potential
field in which particle 1 moves given the actual location of particle
2.  This is, arguably, just what one would expect for the
Schr\"odinger-like equation (for the pilot-wave) of a single particle, 
in a pilot-wave TELB.  

In any case, it is clear (by thinking about the evolution in
configuration space, as indicated in Figure 1)
that entanglement is going to develop when the
pilot-wave fields for the two particles
start to overlap in physical space.  (Entanglement
would develop even earlier if there were long-range forces between the
particles.)  We can see how and when this occurs, from the perspective
of the current theory, by examining Equation \eqref{barschlike}.  The
only term on the right hand side which can produce entanglement when
there is none to begin with is the term involving $\partial V /
\partial x_2$ (evaluated at $x_2 = X_2(t)$ and then again at the full
configuration point) and $\psi_1(x,t)$.  The
factor in parentheses is like the gradient of the ``conditional
potential'' minus that same quantity evaluated at the actual location
of particle 1.  One might call it the ``relative conditional potential
gradient'' -- which would surely not be worth naming in those
cumbersome words, if it weren't for the
fact that this quantity is uniquely responsible for producing
(first-order) entanglement between the particles.  

For the potential involved in our example -- $V(x_1,x_2,t) \sim
\delta(x_1 - x_2)$ -- and given that particle 2 is (at the beginning
of the experiment) more or less stationary near the origin, it follows
that the ``relative conditional potential gradient'' has non-trivial
structure (basically, the derivative of a delta function) also near
the origin.  So this term comes into play precisely when the incoming
particle 1 wave packet, $\psi_1(x,t)$, begins to have support near the
origin.  This produces (for the first time) non-zero values for
$\bar{\psi}_1$, which field then acquires a non-trivial dynamical
evolution according to Equation \eqref{barschlike}.  And, of course, the
entanglement also feeds back into the dynamical evolution of the
pilot-wave field $\psi_1(x,t)$ -- and so also the motion of particle 1
-- through the appropriate terms in Equation \eqref{schlike}.  And of
course higher-order entanglement is simultaneously being produced and
feeding back into the dynamics of $\bar{\psi}_1$ and $\psi_1$.  In
pattern, it is this production and subsequent feedback of entanglement
which allows our TELB (unlike the earlier TELBs mentioned in the
Introduction) to predict strict energy and
momentum conservation during the scattering process.

Of course, the main undesirable feature of the theory
presented in the previous section is still with us:  a (countably)
infinite number of (interacting) fields is still needed to track the
dynamics of the system and reproduce the quantum predictions. 
We have merely re-organized the information
present in the full configuration space wave function so as to make
the specification of initial conditions more natural; but the same
infinite number of fields are still in play.

There is some reason to think, however, that one could achieve
sufficiently accurate predictions by keeping only a finite (and
perhaps reasonably small) number of
the low-order entanglement fields for each particle.  The basis for
this thought is the fact that, at least in the kind of experiment we
have been considering, entanglement is for all practical purposes
(FAPP) a
transitory phenomenon.  It's not \emph{really} transitory, as
evidenced by the existence of two widely-separated packets in the
configuration space at the end of the experiment.  But it's FAPP
transitory in the sense that whichever (configuration space) 
packet fails to contain the \emph{actual} configuration point, quickly
(in particular, as soon as the packets no longer overlap appreciably)
becomes dynamically irrelevant to the motion of the particles.  There
is, in the usual language of pilot-wave theory, an ``effective
collapse'' in the sense that the conditional wave functions acquire
(again) the structure of a single wave packet surrounding the particle
and satisfying (again) an ordinary one-particle Schr\"odinger
equation.  

One can understand this, in terms of the theory being developed here,
as follows:  the entanglement which is present at the end of the
experiment is now spread over many high-order entanglement fields, and
manifests as structure in those fields which is far away from the
spatial region where the associated 
pilot-wave fields have support.  It thus
affects the pilot-wave fields in just the same way that they'd be
affected by an external 
potential with non-trivial structure only where the pilot-wave fields
fail to have support -- which is to say:  it
doesn't affect them at all.  The entanglement, while still in
principle present, has become dynamically irrelevant to the evolution
of the pilot-wave fields and hence also the motion of the particles.  

Remember, too, that what we are calling the ``entanglement fields''
can be thought of as capturing (via something like Taylor expansion)
the ``diagonal'' structure of the full configuration space wave
function.  And so the thought is:  even a finite-order Taylor expansion 
should do a pretty fair job of capturing this structure in the 
immediate vicinity
of the point one is expanding around, which is here the rectangular
axes which emerge from the actual configuration point (shown, for the
initial and final moments of the experiment, as the dotted lines in 
Figure 1).  

The lesson is that, at least in this type of phenomenon, the
entanglement which is actually relevant to the motion of the particles
is \emph{transitory} and captures information pertaining to regions of the
configuration space wave function that is \emph{not too far away} (in the
configuration space) from the actual particle configuration.  This
suggests that we should be able to capture the relevant entanglement
information in a way that is FAPP \emph{accurate
enough} by tracking only a finite (and not too large) number of the
entanglement fields.  That is, it suggests that we could
\emph{truncate} the hierarchy of entanglement fields, keeping only
(say) the several lowest-order entanglement fields for each particle,
and still have an empirically adequate theory.  

It might be worthwhile to undertake numerical simulations to see what
accuracy can be achieved, for something like 
the example illustrated in Figure 1, by
truncating at different orders.  And it might be worth thinking, in
general, about what kinds of situations would most dramatically
distinguish the sort of ``truncated TELB'' contemplated here with ordinary QM.
And it might be worth studying more carefully the question of 
whether the dynamics of such a ``truncated TELB'' is even mathematically
well-defined.  

More likely, though, at least at present, such investigations 
would be simply premature.  The point here is not
really to advocate this particular theory. 
More thinking is needed about how to
best capture the dynamically (most) relevant aspects of the 
entanglement using fields on physical space (or other local beables).  
And more thinking is needed about
whether any such theory could be made to work also in scenarios (such
as those involving bound states rather than scattering) in which
dynamically relevant 
entanglement persists over long periods of time.  And there are also some
other reasons (to be discussed shortly) not to take the present theory
too seriously.  
Our goal here, then, is simply to show that the relevant
entanglement information (necessary to reproduce the strict energy and
momentum conservation that was a specific problem for early ideas in
the TELB direction)
can, in fact, be captured by local beables --  and to suggest
that, in principle, a ``reasonable'' number of additional fields (or
other local beables) might be
sufficient to reproduce, with FAPP sufficient accuracy, the predictions of
ordinary quantum theory.

\section{Discussion}
\label{sec5}

We have shown that, in principle, it is not hard to construct --
starting from extant formulations of pilot-wave theory involving the
configuration space wave function as a non-local beable -- a theory of
exclusively local beables (TELB).  The sort of theory we have
suggested is not \emph{precisely} of the
sort envisioned by many of QM's founders (that is, a theory in which
the usual wave function on $3N$-dimensional space is replaced by $N$
fields on $3$-dimensional space).  Instead, this naively-expected 
ontology forms a foundation on which the full ontology of the new
theory is built:  it \emph{contains} $N$ particles and $N$ pilot-wave
fields on physical space, but includes also a number of \emph{additional} 
fields (also on physical space) which capture what is described in ordinary
QM as ``entanglement'' between particles. 

In principle, at least with the kind of scheme explored here, 
it seems that a (countably) infinite number of these
additional fields is required to \emph{precisely} reproduce the
predictions of ordinary QM.  But as suggested in Section \ref{sec4}
there is
some reason to think that, for practical purposes, a finite number of
such fields could suffice for empirical adequacy.  Of course, this way
of putting the point makes it sound like the purpose of the new theory
is merely to replace ordinary QM with an alternative, in some ways
simpler (and for all practical
purposes accurate enough) computational algorithm.  The perspective to
be considered, though, is just the opposite: 
the world might include just a finite number of these entanglement
fields, in which case some ``truncated pilot-wave theory of
exclusively local beables'' like that sketched in the previous section
would describe reality \emph{exactly}, and it would be the ordinary
quantum theory (with its configuration space wave function) which
would be merely a convenient computational algorithm, accurate enough
for most practical purposes.  

It is interesting to briefly consider the kinds of causation that are
present in, say, the theory we sketched in Section \ref{sec4}.  It is
clear, first of all, that each particle is ``piloted'' by its
associated pilot-wave field.  In turn, the evolution of a
given pilot-wave field is determined by:  (i) the structure of
that field itself, as in the usual, one-particle Schr\"odinger
evolution; (ii) the structure of the associated ``entanglement
fields'' from one and two orders up; and (iii) facts which in some way
pertain to the precise position and/or motion of the other particle.
Included in (iii) are several of the terms on the right 
hand side of Equation \eqref{schlike} which involve, for example, 
the instantaneous
velocity of particle 2 and/or facts about the structure of particle 2's
pilot-wave field in the immediate vicinity of particle 2.  And
of course the potential energy function (which acts like the potential
energy term in the usual, one-particle Schr\"odinger equation) is in
fact the ``conditional potential'', i.e., the potential field in which
particle 1 moves \emph{given the actual location of particle 2}.  

It is sometimes raised as an objection against pilot-wave theory that,
in the theory, the wave function causally influences the particles,
but the particles exert no influences back on the wave. \cite{anandan}
(This, it is
apparently thought, suggests that the particles are some kind of mere
epiphenomenon, which might as well be dropped -- a bizarre
suggestion, for anyone who understands the crucial role the particles
play in making the theory empirically adequate, but still a suggestion
one hears sometimes.)  To whatever extent one takes such an objection
seriously, then, it is of interest to point out its inapplicability to
the pilot-wave theory (of exclusively local beables) sketched here:
each particle's motion is dictated just by its own associated
pilot-wave field, but the evolution of each pilot-wave field is 
influenced by
all the other particles.  Not only, then, do the particles influence
the pilot-wave fields, but the particles can quite reasonably be
understood as (indirectly) affecting each other (through the various
fields).  Perhaps those who dislike the causality
posited by the usual pilot-wave theory, then, will find the theory
sketched here more tolerable.

In elaborating the theory here, we have focused almost exclusively on the
case of two spinless, non-relativistic particles moving in a single
spatial dimension.  As mentioned already, it is relatively
straightforward to generalize everything we've done to the case of $N$
spinless non-relativistic particles moving in $3$-space.
Incorporating spin is definitely a more serious problem for the
specific theory outlined here, because there is no sensible way to
define conditional wave functions for systems of particles with spin.
\cite{au}
One might contemplate, instead of incorporating spin
through the wave function as this is done in the currently-standard 
formulations
of the de Broglie - Bohm theory, introducing spin as a genuine
property of particles along the lines of the ``rigid rotator'' model
presented by Holland \cite{hollandbook}; 
this would seem to allow sensible conditional
wave functions for systems of spinning particles, but at a high price
in terms of elegance.  Probably a better approach is to step back and
consider alternatives to using the conditional wave functions as the
``pilot-wave fields.''  The ``conditional quantum potential''
and the ``conditional velocity field'' would seem to be sensible
candidates here since these, like the conditional wave functions,
could be thought of as local beables (in terms of which the motion of
the associated particles can be easily specified).  The ``conditional
density matrices'' discussed in Ref. \cite{densitymatrix} might also
be worth considering here.

Considerations of relativistic generalization seem to push in a
similar direction.  Already in the case of Galilean boosts in the
non-relativistic theory, one has evidence that the conditional wave
functions cannot be taken as the sort of ``physical scalar fields''
our mathematical formulation makes them look like.  \cite{maudlin}
Of course, one
could still treat the conditional wave functions as scalar fields if
one abandoned even Galilean relativity and adopted an ``Aristotelian''
space-time framework.  \cite{valentini}  But it seems much more
natural, and much more likely to lead to sensible generalizations in
terms of both spin and relativity, to consider alternative pilot-wave
fields (such as those mentioned in the previous paragraph) 
as the basis for a pilot-wave TELB.  

Finally, we mention again the role of random initial conditions and
the emergence of Born rule probabilities for experimental outcomes.
We have suggested that one might be able to increase the simplicity
and plausibility of a theory of exclusively local beables by simply
eliminating the high-order ``entanglement fields'' -- as in the
``truncated'' theory contemplated in the previous section.  Such a
maneuvre, while clearly simplifying the theory of exclusively local
beables, would instead, from the point of view of a mathematically
equivalent configuration-space theory, complicate things
tremendously!  That is, the configuration-space wave function (which
would exactly reproduce the particle trajectories predicted by our
truncated TELB) would
evidently obey some kind of complicated, non-linear dynamics, and many
of the important mathematical properties of the usual pilot-wave
theory (such as ``equivariance'' \cite{au}) would no longer hold.  From the
point of view of Bohmian Mechanics, it might therefore be hard to see 
how the ``truncated theory of local beables'' could be any kind of
positive step -- indeed, how such a theory could genuinely be claimed
to make empirically adequate (approximately Born rule) predictions at
all.  

Here we simply suggest a possible answer.  The non-linearities
contained in a configuration-space theory (mathematically equivalent
to a ``truncated theory of exclusively local beables'') would be, in
principle, similar to the modifications (to standard pilot-wave
theory) contemplated by (for example) 
Bohm and Vigier, which were believed to drive
systems toward Born rule probability distributions as a kind of
dynamical equilibrium. \cite{bohmvigier}  Indeed, it seems that Born
rule probability distributions do emerge naturally at a course-grained
level even within the context of the standard, linear pilot-wave
dynamics.  \cite{valentiniwestman}
This suggests that practically any ``small'' change to the
linear dynamics (such as that contemplated in the truncated 
TELBs) might naturally produce Born rule equilibrium distributions.

If a somewhat more plausible TELB can be constructed (along the lines
of the truncated theory sketched in Section \ref{sec4}, or a new
theory with one of the alternative ontologies suggested several
paragraphs back, or something wholly new) it would be quite
interesting to investigate in a more careful way the nature and extent
of the differences in empirical predictions between standard QM and
the candidate TELB.  Such differences might arise from actual
dynamical inequivalencies (such as those entailed by the truncation
scheme suggested earlier), or inequivalencies pertaining in some way
to the implementation or implications of random initial conditions, 
or from issues arising from the attempt to develop a TELB which
reproduces the predictions of ordinary QM even for non-analytic wave
functions, or from some other novel factor not anticipated here.

For purposes of the present paper, though, our goal is merely to
identify the existence of a large tract of interesting but as-yet
completely uncharted theoretical territory -- not actually to start
charting it in any serious way.  The actual charting will be left for
future research, which it is hoped this paper will help stimulate. 

We close with a final remark about ``interpreting quantum theory'' and
the attempt to discover the real nature of physical systems.  The
earliest period in the development of quantum theory (from Einstein's
paper in 1905 until, say, de Broglie's mature pilot-wave theory of
1927) was relatively healthy from a philosophical point of view.   
But the
subsequent period (after the ascendancy of Bohr's Copenhagen approach)
had, in retrospect, some clearly unscientific elements and was indeed
detrimental to the
ultimate project of understanding the real world (however practical it
might have been in some proscribed, short-term sense to focus on
calculations and temporariliy leave questions of ontology aside).  Largely
due to the positive influences of J.S. Bell, the community seems to be
recovering from its positivistic ways.  To be sure it is still a
minority of physicists who think, for example, that it is important to
worry about which realist ``interpretation of quantum theory'' might
be correct -- though the mere existence of such a (growing) minority
itself represents a major turn-around in recent decades.  

Unfortunately, though, the most dominant approach to trying to uncover
the reality behind quantum mechanics, is to let the theory itself tell
us -- as people sometimes say, to simply ``read the correct ontology
off'' from the equations of the theory.  This approach explains (and its
dominance is best established by) the popularity of so-called ``many
worlds'' versions of quantum mechanics, in which the wave function
itself -- evolving always according to the unitary Schr\"odinger
equation -- is taken as a complete description of physical reality.  

We would like to suggest that this approach is completely flawed,
and in fact quite misleading.  One of the lessons of the current work
is that mathematically equivalent formulations can have radically
different implications in regard to ontology and causality.
(The theory presented in Section \ref{sec3}
contains an infinite number of interacting fields on physical space
and causal influences from particles onto the fields associated with
other particles -- but is mathematically equivalent to standard
pilot-wave theory in which there is just one wave, on configuration
space, and no causation from the particles onto the pilot-wave.)  
So it seems quite naive to
think that one is going to learn anything useful about external reality
by staring at some one particular mathematical formulation of a
theory, and ``reading off'' an ontology in the most straightforward,
naive possible way.  

On what, then, should we base our beliefs about the true ontology of
the external world?  And what is the proper relationship between
ontology and mathematically-formulated physical theories?  Looking at
the history of physics suggests an answer.  For example, people
didn't come to believe in the real existence of electric and magnetic
fields because vector-valued functions on 3-space were present in
Maxwell's equations.  Rather, the belief in a field ontology came
first (via Faraday), and it was only because this ontology was, at
least in a qualitative way, \emph{settled}, that Maxwell was able even
to conceive the project of working out the detailed mathematical laws for
the time-evolution of those fields.  

This same pattern seems to play itself out repeatedly in the history
of physics:
one tries to work out the correct ontology, the correct slate of beables,
\emph{first} -- by making abductive inferences from the qualitative behaviors 
observed in certain key experiments -- and \emph{then} one worries about how 
to formulate/infer the mathematical laws governing the behavior of
those beables.  

This is the approach taken by the early developers
of quantum theory discussed in the Introduction.  It is, for example,
quite explicit in one of the remarks of John Slater quoted earlier.  
And already in his 1905
paper, Einstein is citing certain key experiments as the
basis for the (tentatively) proposed pilot-wave ontology:
\begin{quote}
``It seems to me that
the observations associated with blackbody radiation, fluorescence,
the production of cathode rays by ultraviolet light, and other related
phenomena connected with the emission or transformation of light are
more readily understood if one assumes that the energy of light is
discontinuously distributed in space.''  \cite{einstein}
\end{quote}
Keeping this
approach in mind helps one appreciate the fundamental virtue of the
pilot-wave approach.  It's not that pilot-wave theory solves the
measurement problem (though it does, and this is important), nor that
the pilot-wave theory allows one to dispense with the usual
``measurement axioms'' of orthodox QM and (probably) derive the Born
rule (though it does these, too, and these, too, are important).  It's
rather that the pilot-wave theory provides far and away the simplest,
most natural, and most plausible way of understanding all of the key
experiments which motivated the creation of quantum theory in the
first place.  Einstein and all the other early founders of the theory
perceived this clearly.  J.S. Bell, I think, tried to remind us of it
when he wrote:
\begin{quote}
``While the founding fathers agonized over the question 
\begin{center}
`particle' \emph{or} `wave' 
\end{center}
de Broglie in 1925 proposed the obvious answer
\begin{center}
`particle' \emph{and} `wave'.
\end{center}  
Is it not clear from the smallness of the scintillation on the
screen that we have to do with a particle?  And is it not clear, from
the diffraction and interference patterns, that the motion of the
particle is directed by a wave?  De Broglie showed in detail how the
motion of a particle, passing through just one of two holes in a
screen, could be influenced by waves propagating through both holes.
And so influenced that the particle does not go where the waves cancel
out, but is attracted to where they cooperate.  This idea seems to me
so natural and simple, to resolve the wave-particle dilemma in such a
clear and ordinary way, that it is a great mystery to me that it was
so generally ignored.''  \cite[p 191]{bell}
\end{quote}
While agreeing entirely with Bell here, we also think it is
appropriate to concede that there was at least one puzzling feature of
de Broglie's proposal:  the wave, whose real existence is supposed to
have been inferred from the patterns particles make in 3-space, doesn't
itself live in 3-space.  So in a way de Broglie's theory turned
out not to be about what he intended for it to be about.  

This, we have suggested, was probably the root of (at least) Einstein's
dissatisfaction not only with de Broglie's theory (and later Bohm's),
but the whole ``Schr\"odinger business'' -- i.e., all of quantum
theory.  We have shown, though, that there is actually no significant
roadblock in the way of the kind of theory Einstein (and others)
wanted, i.e., a theory whose ontology matches (or at least includes)
the beables one can
apparently read off from phenomena like the two-slit experiment.  In
fact, there seems to be an abundance of such theories to consider,
develop, and perhaps empirically test.  

It is hoped that the ideas presented here will help remind people of
this apparently long-forgotten, but prematurely-abandoned, approach to
understanding quantum theory and the world it describes.

\section*{Acknowledgements:}
Thanks to Shelly Goldstein, Daniel Victor Tausk, Rob Spekkens, Eric
Dennis, and two anonymous referees for helpful comments  -- and
in particular for a number of very good, constructively critical
questions, many of which remain insufficiently answered.

\end{document}